\documentclass[useAMS,usenatbib,]{mn2e}
\usepackage{graphicx}
\usepackage{subfigure}
\usepackage{epsfig}
\usepackage{mathrsfs,amssymb}
\usepackage{amsmath}
\usepackage{ulem}

\title{The contribution of the IGM and minihalos to the 21 cm signal of reionization}
\author[B. Yue, B. Ciardi, E. Scannapieco \& X. Chen ]{Bin Yue$^{1,2,4}$,
 Benedetta Ciardi$^{2}$, Evan Scannapieco$^{3}$, Xuelei Chen$^{1}$
 \\ $^{1}$National Astronomical Observatories, Chinese Academy of Sciences,
20A Datun Rd, Chaoyang District, Beijing 100012, China
 \\ $^{2}$Max Planck Institute for Astrophysics; Karl Schwarzschild Str. 1;
85741 Garching; Germany
 \\ $^{3}$School of Earth and Space Exploration, Arizona State University,
P.O. Box 871404, Tempe, AZ, 85287-1404
 \\ $^{4}$Graduate University of Chinese Academy of Sciences, 19A,
Yuquan Road, Beijing 100049, China 
 }

\begin{document}

\maketitle

\begin{abstract}
We study the statistical properties of the cosmological 21~cm signal from both the intergalactic
medium (IGM) and minihalos, using a reionization simulation that
 includes a self--consistent treatment of minihalo photoevaporation.  We consider
two models for minihalo formation and three typical thermal states of the IGM 
\--- heating purely by ionization, heating from both ionizing and $\rm Ly\alpha$ photons, and
a maximal "strong heating" model. We find that the signal from the IGM is almost always dominant over 
that from  minihalos. In our calculation, the differential brightness temperature,
${\delta}T_b,$ of minihalos is never larger than 2 mK.
Although there are indeed some differences in the signals from the minihalos and from the IGM,
even with the planned generation of radio telescopes it
will be unfeasible to detect them. However, minihalos significantly affect the
ionization state of the IGM and the corresponding 21~cm flux.
\end{abstract}

\section{INTRODUCTION}

The 21~cm  signal from neutral hydrogen has the potential to provide a
direct and independent probe of the  cosmological gas distribution and
its thermal and ionization states before and during reionization.
Indirectly, such observations can be used to study the formation and
evolution of  early luminous sources, their feedback on the
intergalactic medium (IGM), and their impact  on the formation of
subsequent generations of structures.  As the 21~cm line is associated
with the hyperfine transition in the ground state of neutral hydrogen,
it depends on the distribution of atoms in these two  energy levels,
which is quantified by the spin temperature $T_s$. Since all the atoms
are immersed in a sea of CMB radiation, in principle they can absorb
or emit   photons with the frequency corresponding to the energy
splitting between the two hyperfine levels,
$\nu_0=1.4204\times10^3\rm MHz$, possibly changing the shape of the
CMB  spectrum itself.  If only CMB photons are present, absorption and
emission   equilibrate in a short time, and no line will be visible.
However, there are other two mechanisms that can change the spin
states of neutral hydrogen atoms \--- collisions and $\rm Ly\alpha$
excitation (Wouthuysen-Field effect,
\citealt{Wouthuysen1952,Field1958}), which render the 21~cm line
visible in emission ($T_s > T_{\rm CMB}$) or absorption ($T_s < T_{\rm
CMB}$)  against the CMB background (e.g. \citealt{Madau1997}).

The 21~cm signal from the high redshift universe has been studied in
relation to minihalos (\citealt{Iliev2002,Furlanetto2006a,Shapiro2006}) and the IGM
(e.g. \citealt{Tozzi2000,Ciardi2003,He2004,Furlanetto2004b,Mellema2006,Santos2007}).
Prior to the formation of radiation sources, the IGM is cold and neutral and the 21~cm line is expected
to be observed in absorption against the CMB. Observations of the 21~cm signal at these very high 
redshifts would provide a wealth of information about the initial density fluctuations and the 
presence of a running spectral index or deviations from gaussianity
(\citealt{Loeb2004,Ali2005,Pillepich2007,Lewis2007,Cooray2008,Mao2008}). Once the ionization
sources turn on, they affect the 21~cm signal both by reducing the amount of neutral hydrogen and
by increasing the temperature of the IGM though photoheating, 
$\rm Ly\alpha$ photon scattering and, if the sources are hard enough, X-ray heating 
\citep {Madau1997,Oh2001,Chen2004,Chen2006,Chuzhoy2006,Furlanetto2006c,Chuzhoy2007,Ciardi2007,Semelin2007,
Ripamonti2008}.  Eventually, such heating renders the IGM visible in emission.        

A 21~cm signal is also expected from minihalos (MHs), collapsed
structures with masses larger than the Jeans mass and smaller than the
mass  corresponding to a virial temperature of $\approx 10^4\rm K$.
In such halos, the gas  cannot cool efficiently unless a large
fraction of molecular hydrogen is present. This, together  with their
fragility with respect to feedback effects (see
e.g. \citealt{Ciardi2005}) makes  their contribution to star formation
and photon production small at best.  Nevertheless, MHs are expected
to form copiously throughout cosmic history and their detection would
be an observational breakthrough.  It has been suggested  that a
possible way of detecting MHs is via the 21~cm line
(\citealt{Iliev2002,Furlanetto2006a,Shapiro2006}) .  Because of their
high density and  temperature, MHs result in a signal that is expected
to be different from that of the IGM. On one hand, collisions  are
frequent in such halos, coupling $T_s$ to the gas temperature; on the
other hand, the  very dense gas can trap ionization fronts, protecting
their central regions from being ionized. This makes MHs a possibly
significant or even dominant source of 21~cm emission at some stage of
cosmic history.

As MHs are expected to be found preferentially in high density regions, they are more likely to absorb UV
radiation produced by nearby sources and get progressively photoevaporated
(\citealt {Shapiro1997,Shapiro2000,Haiman2001,Barkana2002,Shapiro2004,Iliev2005,Ahn2007,Whalen2008}). 
As a consequence, they can deplete the number of ionizing photons and delay IGM reionization. In addition to photoionization, MH formation and evolution is affected by a 
variety of feedback effects (see \citealt{Ciardi2005} for a complete review), whose relative 
importance has not been clearly established yet.
\citet[C06 hereafter]{Ciardi2006} performed the first simulations of reionization
with a self-consistent treatment of MH photoevaporation, finding that,
depending on the strength of the feedback effects, the presence of MHs could delay the end
of reionization by as much as $\Delta z \approx 4$.
In this paper, we use the results of these simulations to study the 21~cm signal from both the 
IGM and minihalos. 

The layout of the paper is as follows. In Section 
\ref{methods} we give a short description of the C06 simulations and the method 
used to calculate the 21~cm signal from the IGM and the MHs. In Section \ref{results} 
we present our results and discuss their uncertainties. In Section \ref{con} 
we give our conclusions.  

\section{METHOD}\label{methods}

In this Section we describe the method used to derive the 21~cm signal from the IGM and the MHs.

\subsection{Simulations of reionization with minihalo photoevaporation}

Here we give a short description of the simulation run in C06. Readers 
can find more details in the original paper and references therein. 

The underlying N-body simulation was run in a cubic box of comoving length $L=20h^{-1}\rm Mpc$, 
which was extracted and resimulated from a much larger region 
\citep {Yoshida2001}. Galaxy formation was followed with a semi-analytic model 
\citep {Kauffmann1999} and radiative transfer was calculated with the code {\tt CRASH} 
\citep {Ciardi2001,Maselli2003,Maselli2008}, with the assumption that the gas follows the 
 dark matter. This distribution was then mapped on a $128^3$ grid,
using the Triangular Shaped Cloud interpolation (\citealt {Hockney1981}) to get the 
density in each cell. Throughout the simulation, the \cite {Efstathiou1992} transfer function was used
and the cosmology was taken to be 
$(\Omega_m,\Omega_{\Lambda},\Omega_b,h,\sigma_8,n)=(0.3,0.7,0.04,0.7,0.9,1)$, 
where $\Omega_m$, $\Omega_{\Lambda},$ and $\Omega_b$ are the mean matter,
vacuum, and baryonic densities in units of the critical density, $h$ is the Hubble constant
in units of 100 kms$^{-1}$ \rm Mpc$^{-1}$,  $\sigma_8$ is the rms amplitude of matter fluctuations on 
the 8 Mpc h$^{-1}$ scale, and $n$ is the slope of the primordial power spectrum.

Unlike the original simulations by \cite{Ciardi2003a} and \cite{Ciardi2003b}, in
C06 sub-grid physics was included to take into account the effects of MHs on reionization.
Generally speaking, the box size of a reionization simulation should be large enough to
be representative of the whole universe. This requirement makes it impractical to resolve 
individual minihalos (whose mass may be as small as $10^5$~M$_\odot$).
For this reason, C06 used a ``sub-grid correction'' to include their effects on
reionization, which are mainly due to   two issues:

1) Gas accretion onto MHs, which reduces the density of the IGM;

2) MHs absorption of photons that would otherwise ionize the IGM. Because of their high gas density and recombination rate, 
MHs can consume many more photons than the IGM, relative to their overall mass fraction.

To take into account the above effects, at each step of the simulation in C06, 
the following ``sub-grid correction'' was applied:

\begin{itemize}
\item In each cell, the total mass of MHs was calculated using the Extended Press-Schechter 
(EPS) theory \citep {Lacey1993,Mo1996,Sheth2002}\footnote{A slightly different prescription
is used in cells hosting sources, but their impact is minor.}, and the IGM density was
corrected accordingly.
\item The number of ionizing photons absorbed by MHs in each cell was taken to be:
\begin{equation}
N_{\gamma,\rm MH}=N_{\gamma}(1-\rm e^{-\tau_{\rm MH}}),
\label{Ngam}
\end{equation} 
where $\tau_{\rm MH}$ is the optical depth of MHs, which is proportional to the mass 
fraction of MHs $f_{\rm coll,\rm MH}$, and $N_{\gamma}$ is the number of photons injected into this 
cell.
\item At each step, the fraction of minihalos photoevaporated was computed as:
\begin{equation}
\mathcal F_{\rm MH,eva}=\frac{N_{\gamma,\rm MH}}{n_{\rm H}{\Delta}l^3\bar\xi},
\label{F_mh}
\end{equation}
where $n_{\rm H}$ is the hydrogen number density, ${\Delta}l$ is the size of one cell, 
while $\bar\xi$ is the average number of photons 
consumed by minihalos per total atom (the expressions can be found in the original paper).
\end{itemize}
As mentioned in the introduction, the detailed effect of feedback on MH formation and evolution
is still unknown. For this reason C06 have considered two extreme models in the simulations.
The first is the extreme suppression (ES) model, in which once a cell has been crossed by
an ionizing photon, no new halos are allowed to form within it unless it recombines to become 
completely neutral. The
second is the reformation model (RE), in which MHs continue to form undisturbed by feedback
effects. The real situation is expected to lie between these two extreme cases.

\subsection{21~cm signal from the IGM}
The differential brightness temperature ${\delta}T_b$ of the IGM is calculated 
following \cite {Ciardi2003} 
(see also \citealt {Furlanetto2002, Mellema2006, Shapiro2006}):
\begin{equation}
{\delta}T_b=0.016\mathcal{T}\left(1-\frac{T_{\rm CMB}}{T_s}\right) {\rm K},
\label{dT}
\end{equation}
where
\begin{equation} 
\mathcal{T}=\frac{1}{h}(1+{\delta})(1-f_{\rm coll,MH})(1-x)\left(\frac{{\Omega}_bh^2}{0.02}\right)
\left[\left(\frac{1+z}{10}\right)\left(\frac{0.3}{{\Omega}_m}\right)\right]^{1/2},
\label{T_IGM}
\end{equation}
with $x$ ionization fraction. This 
formula is accurate only in the optically thin limit, which is appropriate for the IGM. $T_s$ 
is the spin temperature, which determines the distribution of neutral hydrogen atoms in the two 
hyperfine levels. The value of $T_s$ is a weighted mean between the CMB temperature, $T_{\rm CMB}$,
and the gas kinetic temperature, $T_k$, and depends on the effect of collisions and 
$\rm Ly\alpha$ pumping. In this work we use 
\begin{equation}
T_s=\frac{T_{CMB}+(y_{{\alpha},eff}+y_c)T_k}{1+y_{{\alpha},eff}+y_c},
\label{Ts}
\end{equation}
where $y_c$ is the collisional coupling efficiency (including H-H, H-e$^-$ and H-p collisions; see
\citealt {Nusser2005,Kuhlen2006}), and $y_{{\alpha},eff}$ is an effective Ly$\alpha$ coupling efficiency term 
that can be written as \citep {Chuzhoy2006}:
\begin{equation}
y_{{\alpha},eff}=y_{\alpha}\exp\left[-0.3(1+z)^{1/2}{T_k}^{-2/3}\right]\left(1+\frac{0.4}{T_k}\right)^{-1},
\label{yalpha}
\end{equation}
where $y_\alpha$ is the standard $\rm Ly\alpha$ coupling efficiency \citep {Madau1997}.
The IGM temperature in the absence of heating mechanisms has been calculated using
{\tt RECFAST} (\citealt{Seager1999}, 2000).

Typically, the IGM can be found in three thermal states, which define the value of the differential
brightness temperature, as listed below:

1) {\it No heating.}  In this state, which occurs before and during the early stages of 
structure evolution, the IGM cools adiabatically.
As long as the density of neutral hydrogen is high enough, collisions between different atoms 
couple $T_s$ to $T_k$, which is usually lower than $T_{\rm CMB}$ at that stage, 
so that the 21~cm signal is in absorption. However, following the expansion of the universe, 
the density decreases as $\propto(1+z)^3$, so that collisional coupling becomes ineffective 
below $z \approx 20$. In this case, the spin temperature is only slightly smaller than
$T_{\rm CMB}$.

2) {\it $\rm Ly{\alpha}$ and UV photon heating}. 
Once the first sources of radiation turn on, their UV photons start to ionize and heat
the surrounding gas. While their impact is limited to the regions in the immediate
vicinity of the sources, the $\rm Ly{\alpha}$ photons (both continuum
and injected; see \citealt{Chen2004,Chen2006}) can
travel much further from their production sites because of the longer mean free path
compared to UV photons, and quickly build up a background that can influence the IGM
as a whole. The $\rm Ly{\alpha}$ photons have a double
effect on the spin temperature. First, even a small $\rm Ly\alpha$ intensity
($\approx 10^{-20}\rm erg~\rm cm^{-2}\rm s^{-1}\rm Hz^{-1}\rm sr^{-1}$) can make $y_{\alpha}$ in 
eq.~(\ref{yalpha}) much larger than 1 and bring $T_s$ very close to $T_k$, so that the
spin temperature traces the state of the IGM. 
This effect is typically referred to as $\rm Ly\alpha$ coupling or the Wouthuysen-Field effect. 

\begin{figure}
\centering{\resizebox{8cm}{8cm}{\includegraphics{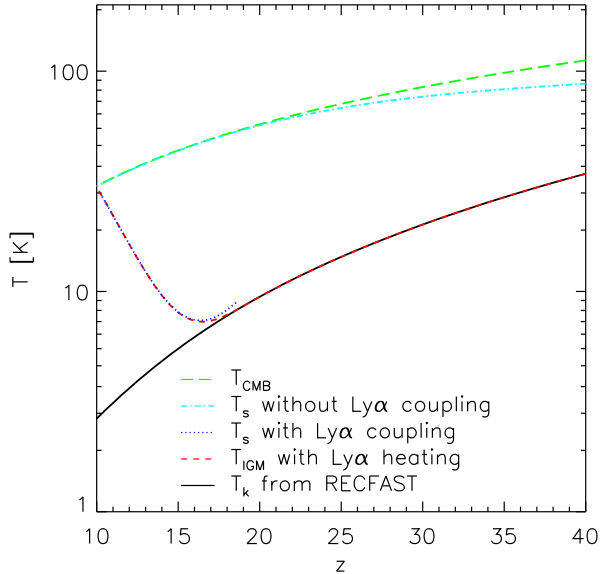}}}
\caption{Evolution of CMB temperature (long-dashed line), IGM temperature in the absence of heating
mechanisms (solid), with $\rm Ly\alpha$ heating (short-dashed), spin temperature without (dashed-dotted line) and with (dotted) $\rm Ly\alpha$ coupling. $\rm Ly\alpha$ photons are included only from the start of the simulation, i.e. from $z=18.5$. }
\label{Fig.tem}
\end{figure}

Secondly, $\rm Ly{\alpha}$ photons can heat or cool the IGM slightly by resonance scattering, 
as has been studied by several authors
\citep {Madau1997,Chen2004,Rybicki2006,Furlanetto2006c}. 
Recent works \citep {Chen2004,Rybicki2006,Furlanetto2006c,Meiksin2006} 
find lower values than originally estimated \citep {Madau1997}. Nevertheless, it may
still have some impact on the thermal evolution of the IGM \citep {Ciardi2007}.
In this paper we calculate this effect 
with the $\rm Ly{\alpha}$ background derived in \cite {Ciardi2003}, which is self-consistent with 
the simulation of C06. Figure \ref{Fig.tem} shows the
temperature evolution of a homogeneous cosmic density field. The 
dashed-dotted line is the spin temperature without any heating mechanisms, while the dotted
line is the spin temperature used in our $\rm Ly{\alpha}$ heating case. Both estimates
exclude the photoionization heating. With the adopted $\rm Ly{\alpha}$ flux background, the
temperature of the IGM is never higher than $T_{\rm CMB}$ before the end of reionization.
Thus, as $\rm Ly{\alpha}$ heating is not extremely efficient, we expect a 21~cm signal 
in absorption. 

3) {\it Strong heating}. Finally, when the $\rm Ly\alpha$ background becomes strong enough or
additional sources of heating are present, we expect the temperature $T_k$ to become much larger
than $T_{\rm CMB}$. Such high temperatures could be reached in the presence of X-rays, 
that can heat the IGM above 1000~K \citep {Chen2004,Chen2006,Furlanetto2006a,Furlanetto2006b}. 
Although there are still some uncertainties on the 
extent of the X-ray background, it seems plausible to have $T_s \gg T_{\rm CMB}$ at later times
during cosmic evolution. 
In this situation, the 21~cm signal is always in emission (\citealt {Zaldarriaga2004,Santos2005}).

In addition to X-rays, Ly$\alpha$ and UV photons, shocks due to structure formation can 
also heat the IGM \citep {Furlanetto2004c,Miniati2004,Shapiro2006}. 
Shock heating may be significant if the gas is still 
cold at high redshift \citep{Furlanetto2006b}. However, \citet{Shapiro2006} showed that 
even taking into account shock heating, the 21~cm signal from MHs is still 
larger than that from the IGM at $z<20$. In fact, shock heating usually is
 efficient around large halos or in high density regions. The same regions however
 are also likely to be ionized by nearby sources. This may suppress the effect of 
heating by shocks. Also, C06 did not consider 
shock heating because they assumed the gas to trace the dark matter distribution.
So this contribution is not considered further in the present work.

The details about the transition between these three configurations are still not clear, mainly 
because of our persistent ignorance on the details associated with the heating mechanism. Here, 
we estimate the 21~cm signal in three heating case: pure UV heating, Ly$\alpha$ and UV heating and 
strong heating. The UV heating is treated as did in the simulation by C06. 

\subsection{21~cm signal from MHs}
In principle, the differential brightness temperature 
of MHs can be derived by replacing $\mathcal{T}$ in eq.~(\ref{T_IGM}) with:
\begin{equation}
\mathcal{T}_{\rm MH}=\frac{1}{h}(1+{\delta})f_{\rm coll,MH}(1-x)\left(\frac{{\Omega}_bh^2}{0.02}\right)
\left[\left(\frac{1+z}{10}\right)\left(\frac{0.3}{{\Omega}_m}\right)\right]^{1/2},
\label{T_mh}
\end{equation}
which is proportional to the mass fraction of MHs.

However, this formula is valid only in the optically-thin case, while the optical depth 
of an individual minihalo is usually large.
In fact, for a halo with mass $10^5$~M$_\odot$, the typical value of the optical depth along 
a line-of-sight across the center of the halo is $\approx 0.5$ \citep {Iliev2002}, and it drops
as the mass increases because of the high $T_s$ and the low absorption coefficient of high mass 
halos (see \citealt{Iliev2002}). 
\cite {Iliev2002} give the more appropriate result of 21~cm flux and ${\delta}T_b$ for an
optically-thick case, assuming that the halo is a Truncated Isothermal Sphere 
(TIS; \citealt {Shapiro1999,Iliev2001}). In this paper we make use of their results. 

Because of their high density and temperature (compared with $T_{\rm CMB}$), individual MHs 
are expected to always produce a 21~cm signal in emission \citep {Shapiro2006}, with a
large ${\delta}T_b$ (in Fig.~1 of \citealt {Iliev2002}, the typical value is several hundred mK
at frequency $\nu_0=1.4204\times10^3~{\rm MHz}$, and even larger after integration along the line profile).
However, it is extremely difficult to resolve individual halos, because their angular size is
usually less than $1^{''}$. For this reason it is customary to estimate, rather than the 
signal from individual halos, the beam-averaged  effective differential brightness 
temperature as (eq.~(6) in \citealt {Iliev2002}):
\begin{equation}
\bar{{\delta}T_b}=\frac{c(1+z)^4}{\nu_0H(z)}
\int_{M_{\rm min}}^{M_{\rm max}}
{\Delta}\nu_{eff}{\delta}T_{b,\nu_0}A\frac{dn}{dM}dM,
\label{dT_mh}
\end{equation}
where ${dn}/{dM}$ is the mass function of minihalos and $H(z)$ is the Hubble parameter.
Here we take the Jeans mass as the lower mass limit, which is
\begin{equation} 
M_{\rm min}=5.7\times10^3\left[\frac{\Omega_mh^2}{0.15}\right]^{-\frac{1}{2}}
\left[\frac{\Omega_hb^2}{0.02}\right]^{-\frac{3}{5}}\left(\frac{1+z}{10}\right)^{\frac{3}{2}},
\label{mmin}
\end{equation}
as in \citet{Shapiro2006}. 
The upper mass limit is 
$M_{\rm max}=2.8\times10^9(1+z)^{-3/2}M_\odot$, which corresponds to $T_{vir}=10^4$~K
in our cosmology. $A=\pi{r^2_{TIS}}$ is the geometric cross section of a halo, where $r_{TIS}$ is the
truncated radius in the TIS model \citep{Iliev2001}; 
$\Delta\nu_{\rm eff}(z)=[(2\pi\mu)^{1/2}\nu_0\sigma_v/c]/(1+z)$ is the redshifted effective line width
and ${\delta}T_{b,\nu_0}$ is the differential brightness temperature of a single halo at the frequency  
$\nu_0$ (the detailed expressions can be found in \citealt{Iliev2002}).

While ${dn}/{dM}$ can be derived from a Press-Schechter approach,  in this 
case, we need also to account for photoevaporation of the MHs. 
While the simulations 
described in Section~\ref{methods} provide the total MHs  mass fraction in each cell at any
given time, we have no information on the distribution of mass amongst the MHs. We thus 
proceed as follows.
In \citet{Iliev2005} it is shown that the mass evolution of an individual halo exposed to
a flux $F_0= F/{\{10^{56}\rm s^{-1}/[4\pi(1 \rm Mpc)^2]\}}$, with an initial mass
$M_0$ at redshift $z_0$, is a function $M=M(M_0,z_0,z,F_0)$.  The function can be inverted to 
derive $M_0=M_0(M,z_0,z,F_0)$, i.e. from the mass of a halo we can get the mass of its progenitor.
Then, at any redshift and in any given cell, the minihalo mass function, including photoevaporation,
can be written as:
\begin{eqnarray}
\frac{dn}{dM}(M,z)=
{\int}^z_{z_{cr}}\frac{d^2n}{dM_0^{'}dz^{'}}(M_{0}^{'},z^{'})\times \nonumber \\
\frac{dM_0^{'}}{dM}(M_0^{'},M,z^{'},z,F_0)dz^{'} \nonumber \\
+\frac{dn}{dM_0}(M_0,z_{cr})\frac{dM_0}{dM}(M_0,M,z_{cr},z,F_0).
\label{newmf}
\end{eqnarray}
Here, $z_{cr}$ is the redshift at which a photon packet crosses the cell for the first time
and $M_0^{'}=M_0(M,z^{'},z,F_0)$. The first term of the equation takes into account the
evolution of the mass function due to all the halos forming between $z_{cr}$ and $z$, while
the second term describes the evolution in the mass function of halos already in place at $z_{cr}$.
Both $d^2n/dM_0dz(M_0,z)$ and $dn/dM_0(M_0,z)$ are calculated with the EPS theory, following
 C06. The equation above is valid in the reformation case, while in the extreme
suppression model, once a cell has been crossed by ionizing photons, no new halo can form.
In this case, the first term of the equation should be set to zero.

\begin{figure}
\centering{\resizebox{8cm}{8cm}{
\includegraphics{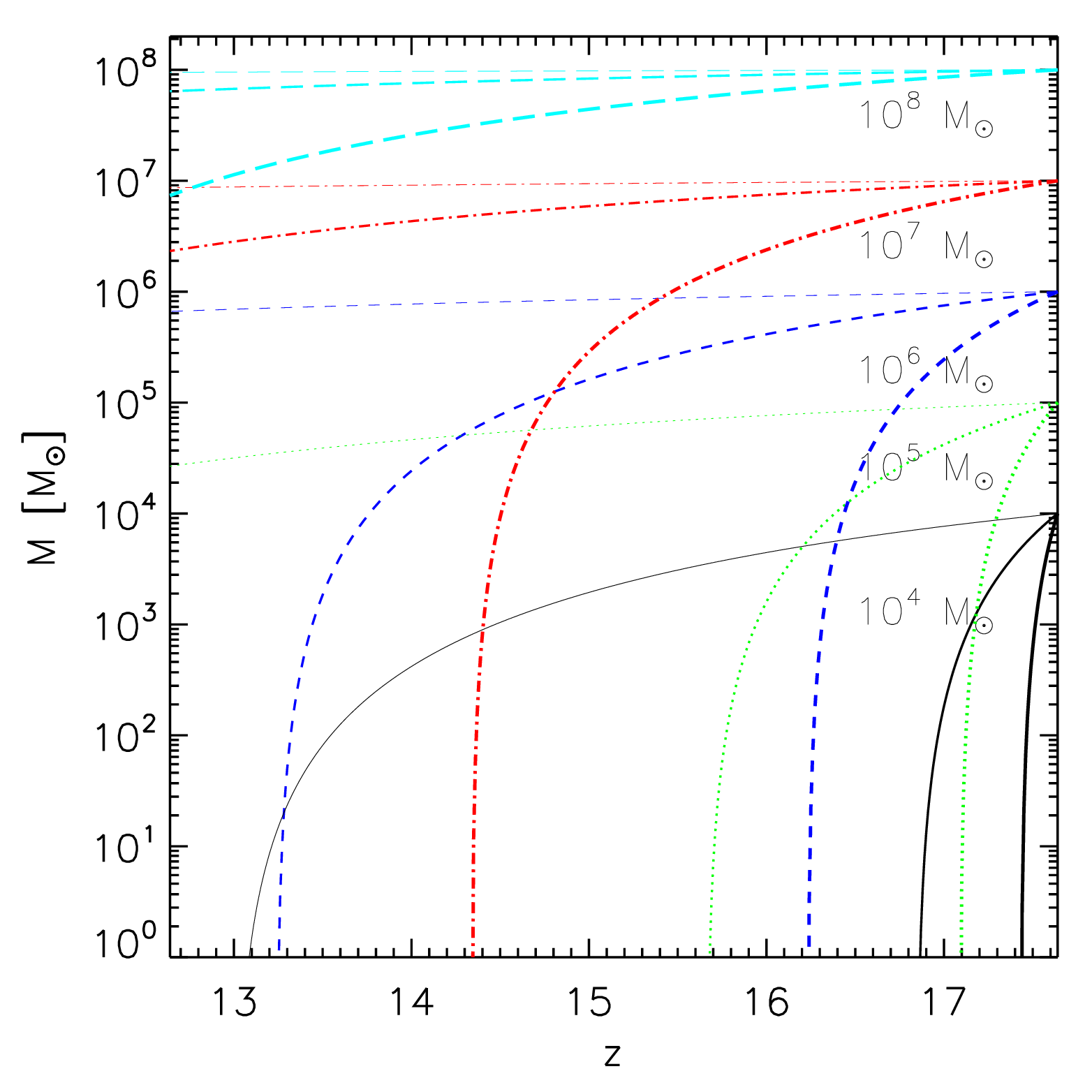}}}
\caption{Redshift evolution of the remaining mass of halos with different initial mass,
under the effect of a photoevaporating flux with $F_0=0.01,1,1000$ (from top to bottom 
in each line group). The initial (at $z_0=17.6$) mass of the halos, 
$M_0$, is as labelled.}
\label{Fig.haloeva}
\end{figure}

Throughout the paper, we use the mass photoevaporation law of \cite {Iliev2005}, which is:
\begin{equation}
M(t)=M_0\left(1-\frac{t}{Ct_{ev}} \right)^B.
\label{mt}
\end{equation}
In this equation, $C$ and $B$ are parameters determined by the spectrum of the sources. In
our case $C=1.07$ and $B=2.5$ \citep{Iliev2005}.  The evaporation time $t_{ev}=t_{ev}(F_0,M_0,z_{cr})$ 
is a function of $F_0, M_0, z_{cr}$  and of the source spectrum. 
Figure \ref{Fig.haloeva} shows the mass evolution of individual halos with different initial masses,
$M_0$, at $z_0=17.6$, exposed to various constant fluxes of 
$F_0$= 0.01, 1, and 1000. While low-mass halos 
disappear on very short time scales ($\approx$14~Myrs if $F_0=1$ and $M_0=10^4$~M$_\odot$), larger mass 
halos retain most of their mass. For example, with $F_0=1$, a halo with $M_0=10^8$~M$_\odot$ 
will have a mass $M=6\times10^7$~M$_\odot$ after 126~Myrs, i.e. has undergone a mass loss of only 40\%.

In principle, if the flux evolution in each cell is known, it is possible
to re-construct the mass evaporation
process for halos in each mass bin according to eq.~(\ref{mt}). In C06, the flux is determined by 
sources in the simulation box \citep{Ciardi2001,Ciardi2006}. However, as
it is computationally unfeasible to record the value of $F_0$ in each cell at each timestep
of the simulation, we only recorded the total evaporated mass fraction of minihalos in each cell.
We then adopt an effective ionization flux which gives identical evaporated mass fraction as $F_0$,
to calculate the mass evolution of halos with different mass. Two methods are used to 
get the effective flux and	implement eq.~(\ref{mt}).
In the first, in each cell and at each time step of the simulation, $z^i$, we use an 
effective ionization flux, $F^i_{0,eff}$, which is assumed to be constant between $z_{cr}$ 
and $z^i$. As we know the total mass of MHs in the cell, 
$M^i_{tot}=\sum_j M_j(t^i)$, we use the following equation:
\begin{equation}
M^i_{tot}=\sum_j M_{0,j} \left(1-\frac{t^i}{Ct_{ev}(F^i_{0,eff},M_{0,j},z_{cr})} \right)^B,
\label{mi}
\end{equation}
where the sum is performed over all the MHs in the cell. The above equation is solved to derive
the only unknown quantity $F^i_{0,eff}$, which is then used in eq.~(\ref{newmf}) 
to calculate the re-distributed MH
mass function and $\bar{{\delta}T_b}$. At the following time step, the same process is repeated
to derive a new effective flux.

In the second method, we also choose an effective flux for each cell. Differently from the first method, 
here the effective flux evolves step by step, rather than being constant from $z_{cr}$ to $z^i$.
At each time step, $z^i$, we calculate:
\begin{equation}
M^{i-1}_{tot}-M^i_{tot}=\sum_j \Delta M^i_j,
\end{equation}
where:
\begin{eqnarray}
{\Delta}M^i_j=M_{0,j}\left[\left(1-\frac{t^{i-1}}{Ct_{ev}(F^{i-1}_{0,eff},M_{0,j},z_{cr})}\right)^B-
\right.\nonumber\\
\left.\left(1-\frac{t^i}{Ct_{ev}(F^i_{0,eff},M_{0,j},z_{cr})}\right)^B\right],
\label{dm}
\end{eqnarray}
is the mass of minihalo $j$ evaporated between two steps. From the above equations we derive
the only unknown quantity, $F^i_{0,eff}$ and then we use it in eq.~(\ref{newmf})
to calculate the re-distributed MH mass function and $\bar{{\delta}T_b}$.
Also, the same process is repeated to derive a new effective flux
at the following time step.
We have used both methods and found that the results are consistent with each other,
as in the vast majority of cells at any time the difference in the ${\delta}T_b$
obtained with the two methods is less than 5\%.
Thus, in the following we will always use the first one.

\begin{figure}
\centering{
\resizebox{8cm}{8cm}{\includegraphics{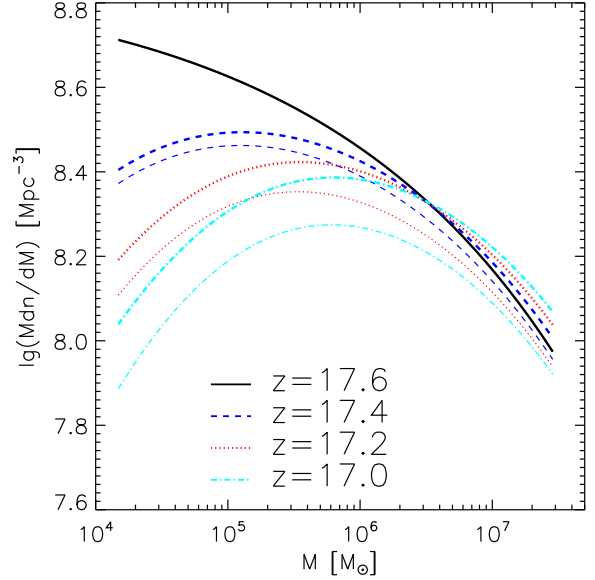}}}
\caption{The evolution of the mass function under photoevaporation for the extreme suppression model
(thin lines) and the reformation model (thick lines). From top to bottom, the lines correspond to
redshift 17.6 (when the evaporation begins), 17.4, 17.2 and 17.0 respectively. Here we assume a constant
flux $F_0=1$.}
\label{Fig.newmf}
\end{figure}

Figure \ref{Fig.newmf} instead shows the mass function at different redshifts with the
assumption of a constant ionization flux $F_0=1$, as calculated in eq.~(\ref{newmf}).
The thin lines refer to the extreme suppression model, while the thick lines are the
reformation model. As expected, the mass function 
in the reformation model is always larger than that in the extreme suppression model. 
In the former, at high mass end, the formation rate of halos is higher than the 
evaporation rate, so the mass function increases with time.
 
\begin{figure}
\centering{\resizebox{8cm}{8cm}{\includegraphics{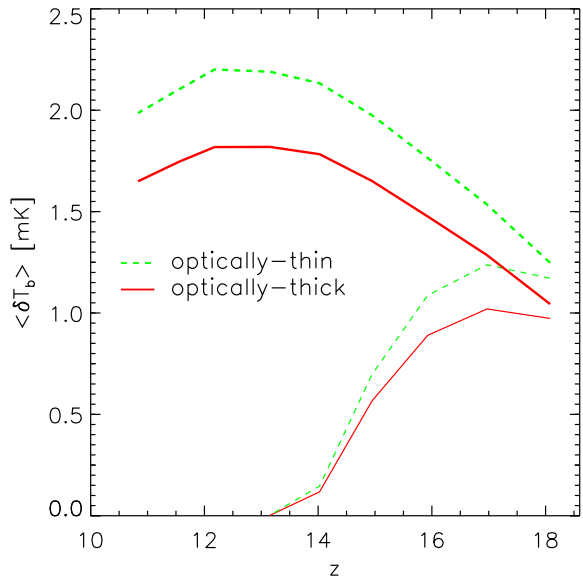}}}
\caption{Evolution of the volume-averaged ${\delta}T_b$ of minihalos from the simulations
described in \citet{Ciardi2006}. The solid (dashed) lines refer to an optically
thick (thin) case. The thin (thick) lines refer to the extreme suppression
(reformation) model for minihalo evolution.}
\label{Fig.dT_mh}
\end{figure}

Using the above mass functions, we can now calculate the expected differential brightness temperature. 
Figure \ref{Fig.dT_mh} shows the mean (volume-averaged) ${\delta}T_b$ evolution for the two MH formation models,
which has been calculated both with the optically-thin formula given in eq.~(\ref{dT}) with the
assumption that $T_s \gg T_{\rm CMB}$ (dashed lines) and for the optically-thick case 
in eq.~(\ref{dT_mh}) (solid lines), with $\delta T_{b,\nu_0}$ taken from Figure~1 of \citet{Iliev2002}. 
Although the ${\delta}T_b$ for an individual halo is very high compared with the optically-thin case,
 the beam-averaged $\bar{{\delta}T_b}$ 
is very small (with a typical value of $\approx 1-2$ mK). This is because the halos usually have a
small geometric cross section $A$, and thus the associated flux is not very significant. 
Figure \ref{Fig.dT_mh} also shows that the optically thin assumption with $T_s \gg T_{\rm CMB}$ is a good 
approximation to calculate the mean 21~cm signal from minihalos.
$\delta{T_b}$ in the extreme suppression model drops dramatically following the evaporation of the 
minihalos. In the reformation model instead, it increases continuously until late times. 
Thus, the presence or absence of MH reformation  can result in  totally different pictures
 of the evolution of the 21~cm signal from MHs.

\section{RESULTS AND ANALYSIS}\label{results}

\begin{figure*}
\centering{\resizebox{16cm}{12cm}{
\includegraphics{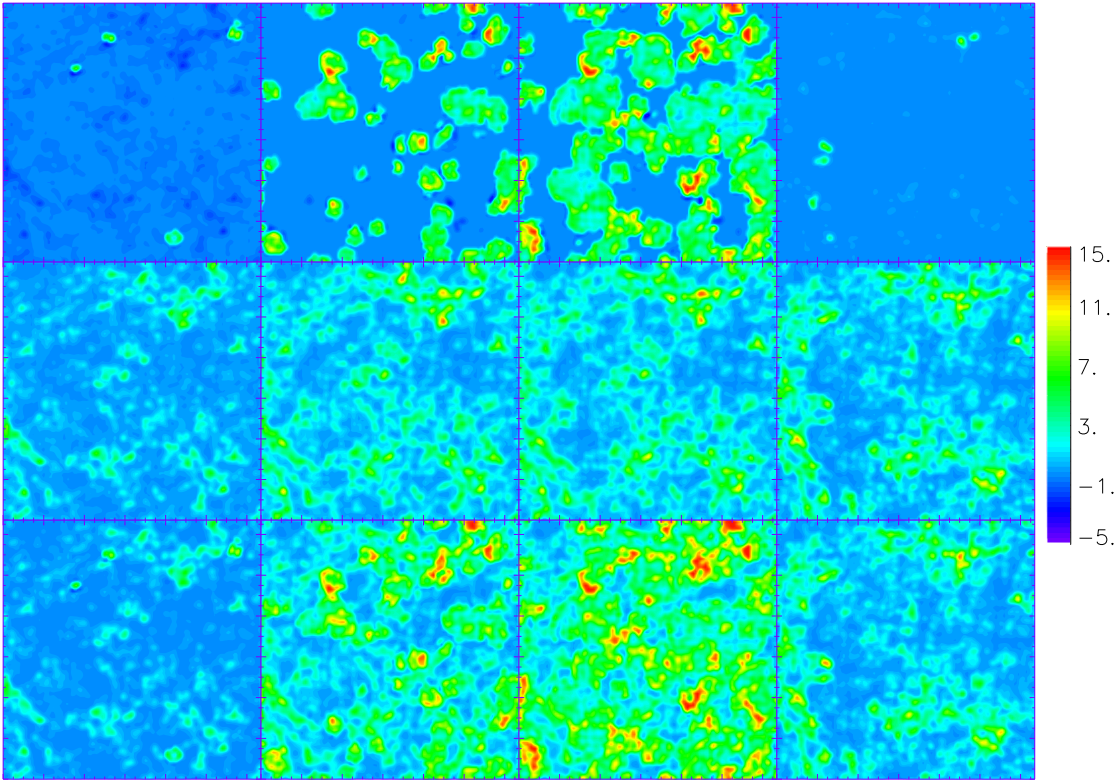}}}
\caption{A map of ${\delta}T_b$ in units of mK in the pure UV heating case and 
reformation case (see text for details). From top to bottom, the three rows correspond to the IGM, 
MHs, and the total. In each row, from left to right, the panels show slices at redshift 18.5,
14.6, 12.8, 10.6 respectively. The comoving thickness of the  slice corresponds to the dimension
of one cell in the simulation, i.e. $\approx 223$~kpc.}
\label{map}
\end{figure*}

In this Section we present our results in terms of mean differential brightness temperature 
(as described in Secs.~2.2 and~2.3) and
fluctuations of brightness temperature, and we discuss their uncertainties.
As a visual example, in Figure~\ref{map} we show maps of the ${\delta}T_b$ in the pure UV
heating case and 
reformation model. Form top to bottom, the three rows correspond to the IGM, 
MHs, and the total.  The map shows that at the early
and late stages of reionization, MHs contribute more to the total signal compared 
with the IGM, while at intermediate redshifts the contribution from 
partially ionized IGM increases.

\subsection{Mean differential brightness temperature}
\begin{figure}
\centering{\resizebox{9.5cm}{15cm}{\includegraphics{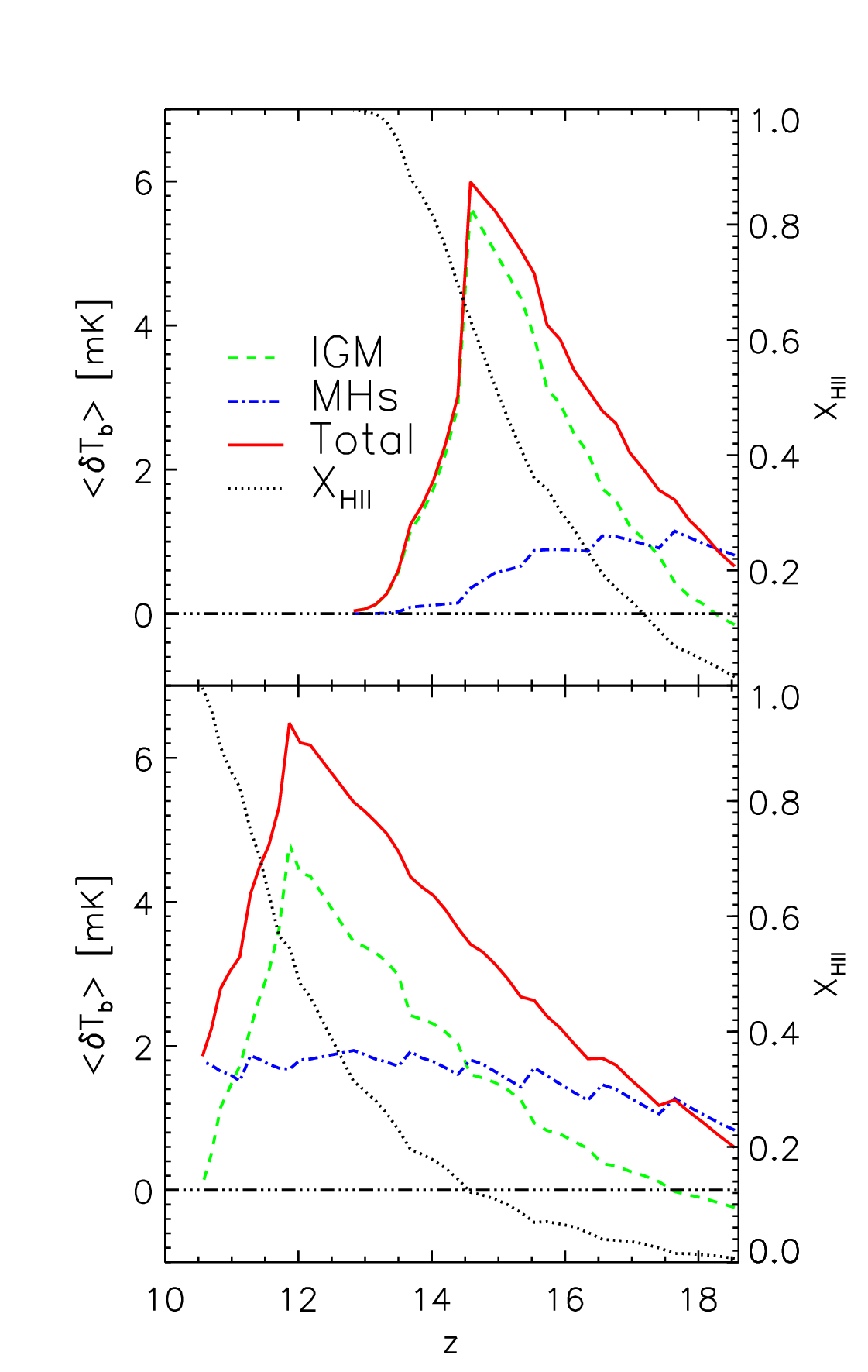}}} 
\caption{Mean ${\delta}T_b$ evolution in the absence of heating sources,
with the exception of the UV ionizing
radiation. The solid, dashed, dashed-dotted and dotted lines refer respectively to
the contribution of IGM+MHs, only IGM, only MHs and the volume averaged ionization fraction.
The top panel refers to the extreme suppression model, while the bottom panel refers to reformation model. 
To guide the eye, we also plot a horizontal line corresponding to ${\delta}T_b=0$.}
\label{Fig.dT_nh}
\end{figure}

The contribution from MHs to the mean differential brightness temperature
is calculated as described in the previous Section and is shown as dashed-dotted lines
in Figure~\ref{Fig.dT_nh}.
The two formation models for MHs (ES and RE) represent two extreme effects of the IGM heating 
on the 21~cm signal from MHs. 
In each model, the signal of MHs does not vary as a function of IGM heating, so these lines are the 
same as the solid lines in Figure~\ref{Fig.dT_mh}.
${\delta}T_b$ from MHs is roughly proportional to the total
mass of MHs, while it is not sensitive to the shape of their mass function.
On the other hand, the IGM contribution strongly depends on its thermal state, and we present
results for three different cases.

First, we discuss the case in which only heating from ionizing radiation is present. 
Over the redshift range covered by the simulations, the collisional coupling in the mean-density IGM 
considered here is very
weak. Thus, in the absence of other decoupling mechanisms, the spin temperature of neutral 
hydrogen is very close to $T_{\rm CMB}$ and the 21~cm line is only slightly visible in
absorption (dashed lines in Fig.~\ref{Fig.dT_nh}). 
However, as the first sources of radiation turn on and reionization begins, 
the energy of the ionizing photons heats the IGM in the vicinity of the sources
to a significant high temperature (in this work we follow C06 and use the reference value $10^4$K, 
which is a reasonable approximation in regions ionized by stellar type sources; see 
\citealt {Ciardi2001,Pritchard2007}).
Thus the IGM in such regions may emit 21~cm radiation until it has been fully ionized, so that 
the mean ${\delta}T_b$ changes from a negative to a positive value. However, as reionization 
proceeds, the amount of neutral hydrogen decreases and the mean ${\delta}T_b$ drops to zero again.

In the extreme suppression case, we find that even in the presence of only collisional decoupling 
and UV heating, the 21~cm signal 
from minihalos dominates the total signal only in the very early stages of reionization.
This is because reionization suppresses the formation of new minihalos and 
photoevaporates existing ones, reducing their contribution with time.
On the other hand, reionization increases the temperature of the partially-ionized IGM and 
the density of free electrons. In high temperature, partially ionized regions, 
H-$e^-$ and H-p collisions are more effective than H-H collisions in coupling $T_s$ to 
$T_k$. As a consequence, the fraction of flux from partially ionized gas is higher than its 
volume filling factor, 
and the signal from the IGM increases with time until a volume averaged ionized 
fraction of $x_{\rm HII} \approx 0.5$ 
has been reached, at $z \approx 15$, which corresponds to a peak of the emission.

On the other hand, in the reformation model, the flux from the IGM is 
reduced by the significant value of $f_{\rm coll, MH}$ and pushed towards lower
redshift by the delay in the reionization process. In this case, the period during which the 
flux from minihalos is larger than that from the IGM is extended to ${\Delta}z \approx 4$, and
 reaches a peak at $z \approx 12.$ 
 
It should be stressed that the signal from the IGM dominates over that from the MHs because
of the collisions between atoms and free electrons and protons (we discuss the 
uncertainties of this in Sec. \ref{uncer}). In fact, if we ignore this
contribution, the peak of ${\delta}T_b$ from the IGM is only $\approx 2$~mK and the signal
from MHs dominates down to $z \approx 15.5$ in the extreme suppression model, and at almost
all redshifts  in the reformation model.

\begin{figure}
\centering{\resizebox{9.5cm}{15cm}{\includegraphics{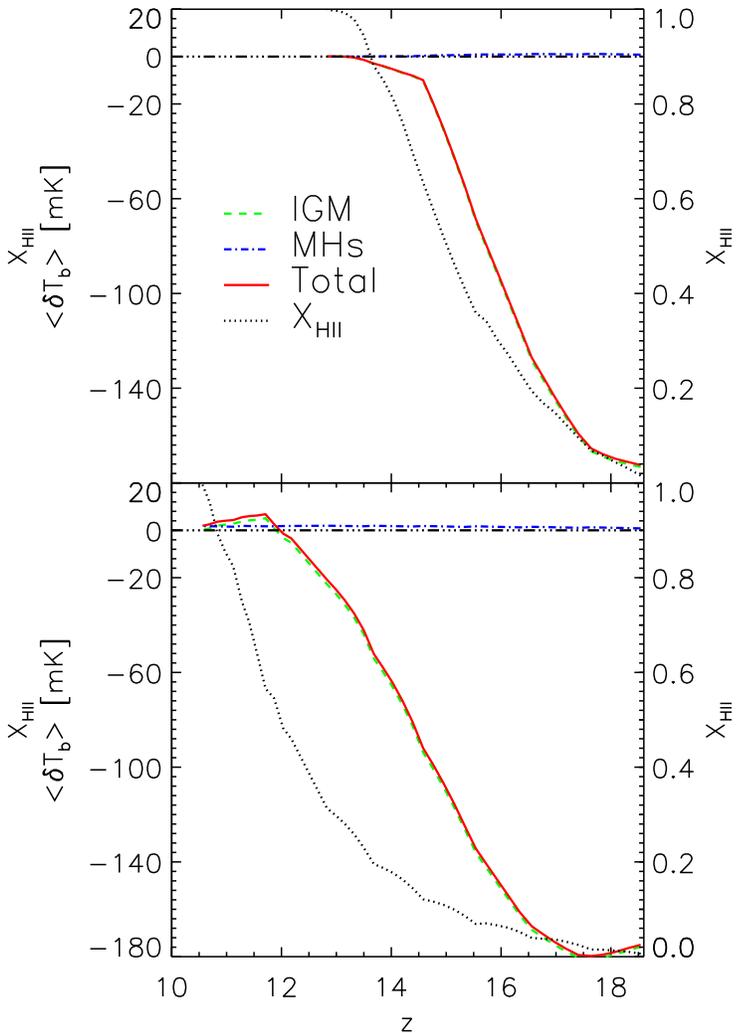}}}
\caption{As Fig.~\ref{Fig.dT_nh} but including heating by $\rm Ly{\alpha}$ photons.}
\label{Fig.dT_hc}
\end{figure}

Next, we consider a case that includes the effects from a $\rm Ly\alpha$ background,
leading to the mean ${\delta}T_b$ evolution shown in Figure \ref{Fig.dT_hc}.  Here we 
see that the $\rm Ly\alpha$ background produced by the sources in the simulation is strong
enough to couple $T_s$ to $T_k$ and to heat the IGM above the adiabatic temperature,
but the mean temperature is still below $T_{\rm CMB}$. This induces a strong mean signal
in absorption, and a minimal emission is visible only at a late evolutionary stage in
the reformation model. This is because reionization proceeds to lower redshift, when
more sources form and produce a larger value of the $\rm Ly\alpha$ background.
Yet even in this case the bulk of the emission  is due to the partially-ionized regions.

The reason for the very different behavior compared to the no $\rm Ly\alpha$ background
case is the following. 
Without $\rm Ly\alpha$ photons the contribution to ${\delta}T_b$ consists of a small
negative signal from the neutral, cold IGM (because $T_s \approx T_{\rm CMB}$),
and a positive signal from the  the partially ionized, hot IGM.
In the presence of $\rm Ly\alpha$ photons instead, the dominant
contribution comes from a strong negative signal from the
the neutral, warm IGM (because $T_s \ll T_{\rm CMB}$) which exceeds
the smaller positive signal from the partially ionized, hot IGM.
In this case we find that the IGM signal always dominates the total signal consistent with
\cite {Furlanetto2006a}.

\begin{figure}
\centering{\resizebox{9.5cm}{15cm}{\includegraphics{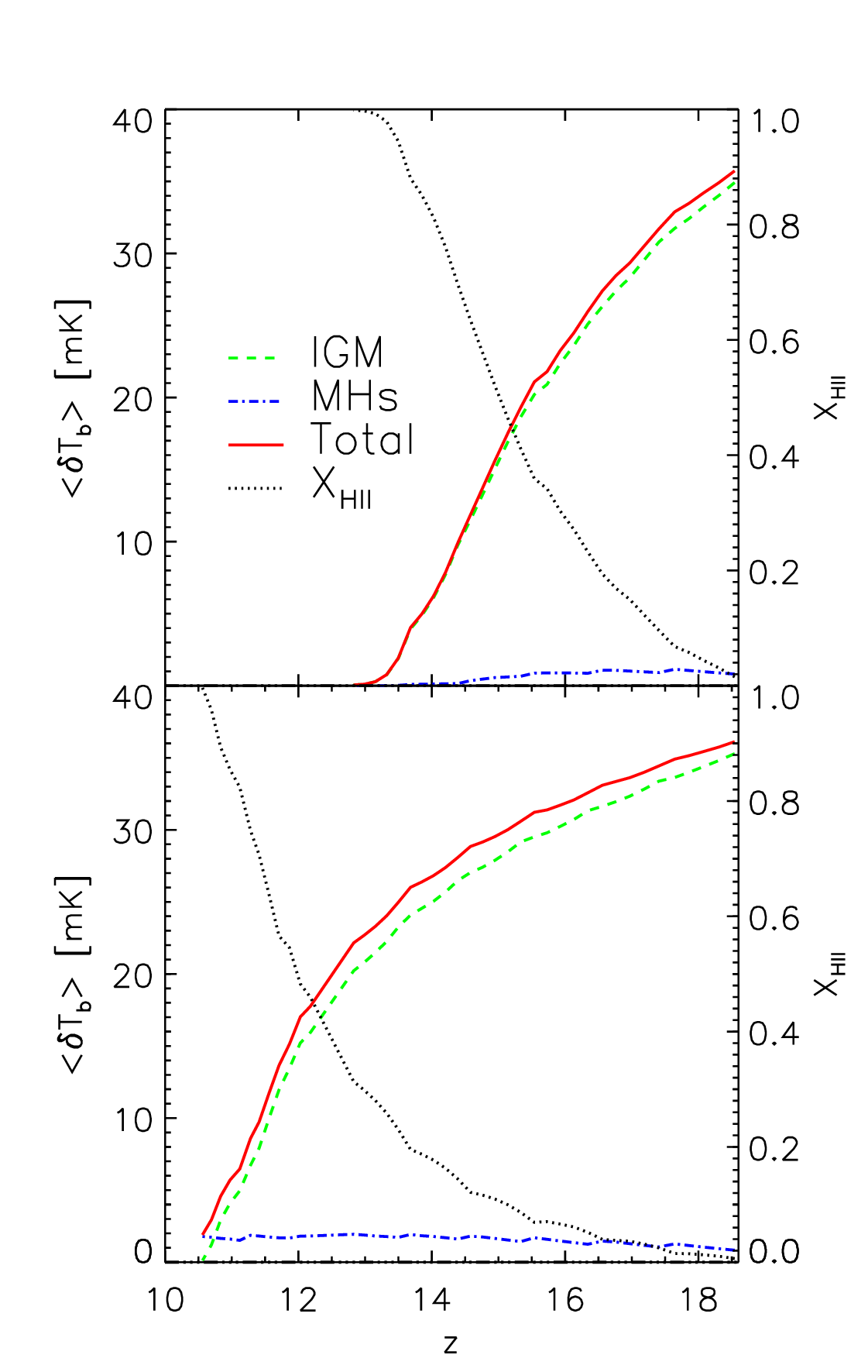}}}
\caption{As Fig.~\ref{Fig.dT_nh} but in the $T_s \gg T_{\rm CMB}$ case.}
\label{Fig.dT_er}
\end{figure}

Finally, we consider a situation in which $T_s \gg T_{\rm CMB},$ resulting in
the evolution of the mean ${\delta}T_b$ shown in Figure~\ref{Fig.dT_er}. 
As expected, in this case ${\delta}T_b$ is always positive and it decreases as reionization
proceeds and the available neutral hydrogen is depleted.
Unlike the cases analyzed before, the signal corresponding to the extreme suppression
model is always lower than that of the reformation model. This is due to the fact that, in
this case, ${\delta}T_b$ is independent of $T_s$ and depends only on the amount of neutral 
hydrogen present at each redshift, which is higher in the reformation model.
Moreover, the signal from the IGM always dominates over that from the MHs. This is because in
both cases the condition $T_s \gg T_{\rm CMB}$ is satisfied, meaning that  
${\delta}T_{b,\rm IGM}\propto f_{\rm IGM,HI} $ and  ${\delta}T_{b,\rm MH}\propto f_{\rm MH,HI} ,$ 
 with a similar constant of proportionality.  Thus $f_{\rm IGM,HI} \gg f_{\rm MH,HI} $ implies
${\delta}T_{b,\rm IGM} \gg {\delta}T_{b,\rm MH}.$ 

\subsection{rms fluctuations and pixel distribution}
We now calculate the rms fluctuations in the brightness temperature distribution,
$\langle({\delta}T_b-\langle{\delta}T_b\rangle)^2\rangle^{1/2},$ as a function of the 
beamsize, ${\Delta}{\theta}$,
and redshift. We choose a slice corresponding to a bandwidth ${\Delta}\nu=0.1$~MHz, 
located at the center of the simulation box. A larger bandwidth (e.g. 1~MHz) would produce 
lower fluctuations because in that case inhomogeneities in the gas density and in the hydrogen ionized 
fractions would be more poorly resolved (see e.g. \citealt {Ciardi2003}).

\begin{figure*}
\centering{\resizebox{12cm}{8cm}{\includegraphics{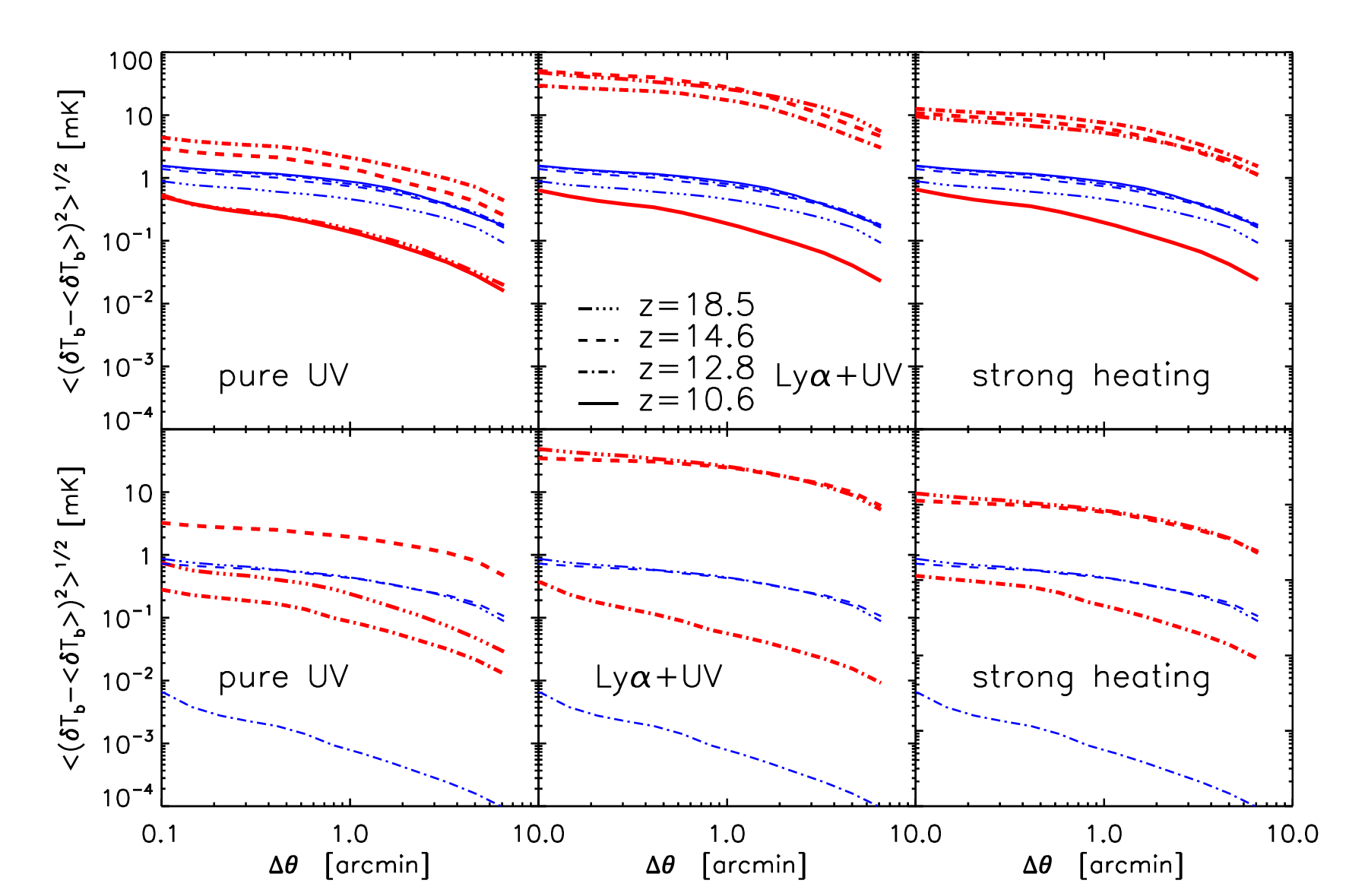}}}
\caption{Expected rms brightness temperature fluctuations as a function of beamsize at
different redshift.
Upper and lower panels refer to the reformation and extreme suppression models, 
respectively. The three thermal states of the IGM
are labeled in each panel. Thick (thin) lines refer to the IGM (MHs) contribution,
at the redshifts 18.6, 14.6, 12.8, and 10.6.}
\label{Fig.rms_theta}
\end{figure*}

\begin{figure*}
\centering{\resizebox{12cm}{8cm}{\includegraphics{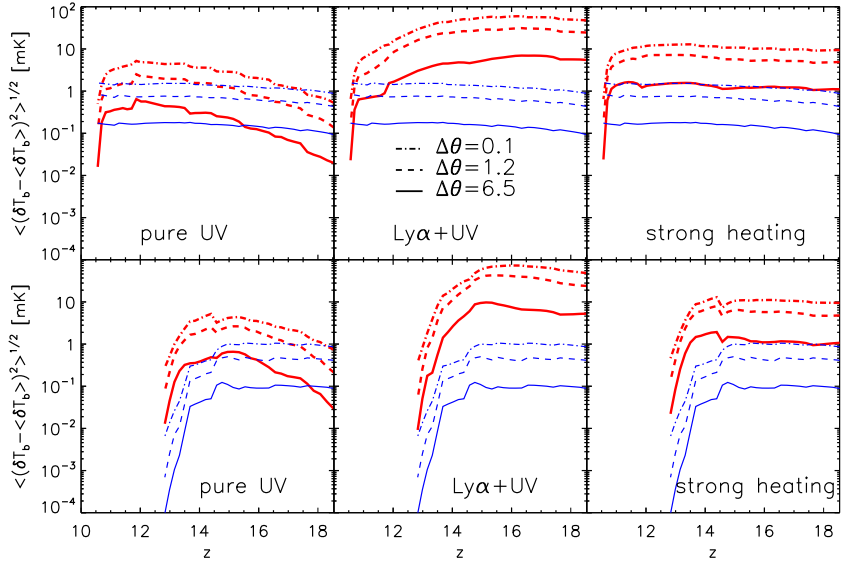}}}
\caption{Expected rms brightness temperature fluctuations as a function of redshift
with different beamsize. The meaning of each panel is the same as in Fig.~\ref{Fig.rms_theta}.}
\label{Fig.rms_z}
\end{figure*}

The results are shown in Figure~\ref{Fig.rms_theta} and \ref{Fig.rms_z},
for the three IGM thermal states and two MH formation models described in the previous Sections. 
Figure~\ref{Fig.rms_theta} gives the rms as a function of 
beamsize at several different redshifts for both the reformation (upper panels) and the
extreme suppression (bottom panels) model.  Figure~\ref{Fig.rms_z}, instead, gives the evolution of 
rms for different beamsize in the same two models. 
In all cases, the signal increases as the angular scale decreases because the mass variance
is larger on small scales. Another common feature is that, although the continuous formation
of minihalos increases the signal compared to the extreme suppression case, it does not 
increase the inhomogeneity of the gas, so the rms in the reformation model is similar to the 
rms of the extreme suppression model until the latter drops because of reionization. 

In the pure UV heating case, the signal from the IGM reaches a maximum at $z \approx 14$ 
(extreme suppression model) and at $z \approx 12$ (reformation model), 
corresponding to the epoch when several neutral regions are 
still present, but the volume is roughly half ionized. Then it starts decreasing and drops 
dramatically when the IGM gets close to complete reionization. The signal from the MHs shows 
a similar behavior, but in the reformation model the drop associated with photoevaporation is 
not evident because they are allowed to continuously reform. The general trend in the relative 
importance of the IGM and MHs contribution follows what already seen in the mean brightness 
temperature, i.e. the MHs contribution dominates in both models in the earliest stages of the 
evolution and when reionization is almost complete in the reformation model. 
Otherwise the IGM contribution is dominant. In addition, the smaller the beamsize, the earlier the 
fluctuations that from IGM exceed those from the MHs\footnote{Here we only 
consider $\Delta \theta= 0.1^{'}$ because this corresponds to the size of one cell in the simulation.}. 
Thus, increasing the resolution of the beam might not help in detecting the signal from MHs
(because observations at higher redshift and lower frequencies are more difficult), 
unless one chooses a region particularly rich in MHs.

Also in the $T_s \gg T_{\rm CMB}$ case and in the Ly$\alpha$ case, the behavior of the rms 
is similar to that of the
mean brightness temperature, with the MH signal dominating only in the reformation model during the
latest stages of reionization.

\begin{figure}
\centering{\resizebox{8cm}{8cm}{\includegraphics{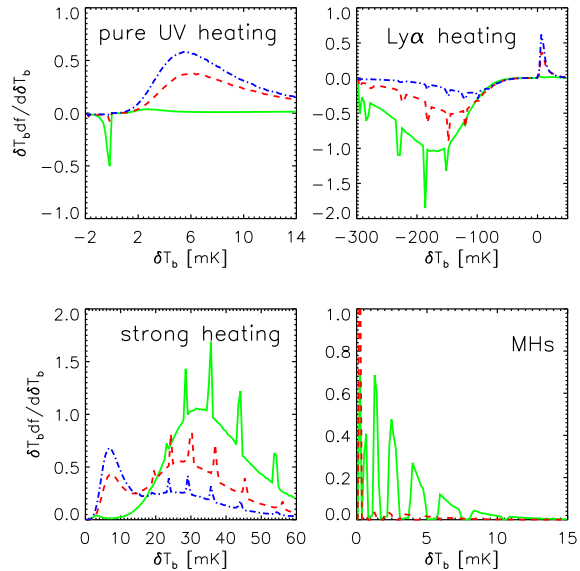}}}
\caption{Pixel distribution for both the IGM and minihalos for the extreme suppression model. 
Different lines correspond to different redshifts. 
In each IGM panel, solid, dashed and dashed-dotted lines mean redshift 18.1, 15.5,
14.8 respectively. In the minihalos panel instead, pixel distribution is given at redshift 
16.6 (solid) and 13.7 (dashed).} 
\label{Fig.pixel_er}
\end{figure}

\begin{figure}
\centering{\resizebox{8cm}{8cm}{\includegraphics{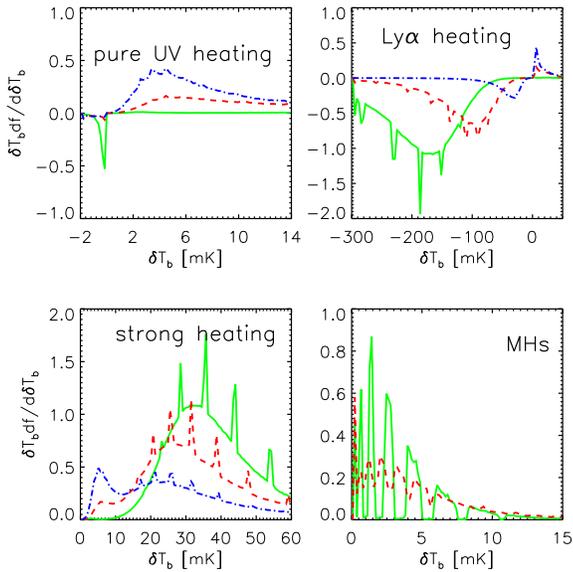}}}
\caption{Same as Fig.~\ref{Fig.pixel_er}, but for the reformation model.
In each IGM panel, solid, dashed and dashed-dotted lines mean redshift 18.1, 14.2
and 12.2 respectively, 
while in the minihalos panel, solid line refers to redshift 16.6 and dashed line refers to 
redshift 11.1.} 
\label{Fig.pixel_f}
\end{figure}

We then calculate the pixel distribution of the brightness temperature,
${\delta}T_bdf/d{\delta}T_b$, as a function of ${\delta}T_b$ for the IGM in the 
three cases and minihalos. Here $df=N_{{\delta}T_b}/N_{tot}$, where $N_{tot}$ is the total number of 
cells in the simulation box and $N_{{\delta}T_b}$ is the number of cells that lie in the 
bin $d{\delta}T_b$. In Figure \ref{Fig.pixel_er} and \ref{Fig.pixel_f} the panels correspond, 
from left to right and from 
top to bottom, to pure UV heating case, Ly${\alpha}$ heating case, $T_s \gg T_{CMB}$ case 
and minihalos, respectively.

The amplitude of the pixel distribution indicates the relative contributions from different 
${\delta}T_b$ bins to the total signal. The double peaks associated with the two-phase
structure of the IGM (totally neutral and partially ionized regions) are very prominent 
(see also \citealt {Ciardi2003}).
For example, in the Ly$\alpha$ heating case, at an early stage (dashed lines), 
when most of the signal comes
from completely neutral regions, only one peak is visible in absorption,
corresponding to ${\delta}T_{b,peak}\approx-150$~mK.
However, as ionization proceeds, the amplitude of 
the peak decreases and moves toward a positive value;
meanwhile, a new, smaller peak appears with positive value at ${\delta}T_b \sim$ several mK, which
comes from the partially ionized regions (see the dashed and dashed-dotted lines in Fig.~\ref{Fig.pixel_er}
and Fig.~\ref{Fig.pixel_f}).
The three redshifts chosen show three typical situations:
when the totally neutral regions dominate, when the partially ionized regions begin to 
produce a significant signal and
when the contribution from partially ionized regions is more evident, respectively. 
The behavior of the pixel distribution is very similar for the extreme suppression and the reformation model,
the only difference being a delay in the reformation model. 
In the $T_s \gg T_{\rm CMB}$ case, the trend is the same as
in the Ly$\alpha$ heating case, i.e. as ionization proceeds, the peak corresponding to the totally 
neutral regions decreases and moves toward zero, 
while the peak originating from the partially ionized regions increases.  
However, the situation is slightly different in the pure UV heating case, in which the contributions from  
the partially ionized regions become significant at an earlier stage.

Finally, the dominant minihalo contribution at all redshifts comes from cells with 
${\delta}T_b\approx$~1-2~mK, which is a typical value of beam-averaged differential brightness 
temperature in our simulation (this value depends on the cosmological parameters and the 
matter power spectrum, see \citealt{Iliev2002}).
Unlike the IGM signal, here there is always only one peak; 
however, since we can measure only a total flux, this feature is not useful to distinguish
the signal of minihalos from that of the IGM. The sawtooth features come from the discreteness 
effect introduced when calculating $f_{\rm MH}$, which shows also in the spikes of the IGM signal.

\subsection{Model uncertainties}\label{uncer}
In this Section we discuss the main approximations and uncertainties of our model and their
impact on the results, in addition to those already discussed in C06.

\begin{itemize}

\item In the pure UV heating case, partially ionized regions contribute significantly to
the emission signal. However, we note that the resolution of our simulation is only
$\approx$~223~kpc, which is not enough to resolve the I-front. This leads to the possibility of
overestimating the signal from partially ionized regions by mixing together totally ionized and totally
neutral regions. In fact there are many uncertainties related to the sub-grid neutral
filling factor, including the inhomogeneity of the density field on small
scales and the spectrum of the sources \citep[e.g.][]{Iliev2006}.

In our work we refer to "partially ionized" cells as regions heated by UV photons that still contain
enough neutral hydrogen to emit a 21~cm signal. Note that this somewhat arbitrary definition
is used only to estimate the relative contribution of gas in different physical conditions,
while all the cells in the simulation have been used to calculate the total 21~cm signal.
In all our previous calculations, we identified partially ionized cells as those with ionization
fraction $x$ above a threshold value of 0.1, but a value of 0.01 leads to similar results.
If $x$ is very close to 1.0, the signal from a cell is negligible and the cell is referred to
as totally ionized.

If we assume that in  a partially ionized cell only a fraction $f_{pi}$ of the gas is
partially ionized with an ionization fraction of $x_{pi}$, while the rest is either neutral and
cold, or ionized and hot and thus irrelevant for the contribution
from the partially ionized regions.   Furthermore, we should consider that only a fraction of gas
$f_{pi} \times (1-x_{pi})$ is contributing to the signal. Both $f_{pi}$ and $x_{pi}$ are unknown,
and vary in a large range cell by cell. As a reference, if we assume $x_{pi}=x$ we find that
at the redshift when half of the gas has been ionized, the mean $\delta T_b$ of the total simulation
volume for $f_{pi}$ equal to (0.0, 0.001, 0.01, 0.05, 1.0) are 
(-0.054, -0.051, -0.012, 0.21, 5.63)~mK respectively 
for the extreme suppression case.  Note that $(f_{pi},\delta T_b)=(1.0,5.63~mK)$ corresponds to the peak of
 the dashed line in upper panel of Fig. \ref{Fig.dT_nh}.

So if the sub-grid neutral filling factor is much lower than 1.0, we expect no emission mean signal
or non-dominant partially ionized IGM signal in the pure UV heating case.
However, the pure UV heating model ignores the pre-heating by X-rays and Ly$\alpha$ photons, thus
we expect it to be appropriate to describe only the early stages of reionization.
Also the small emission in the $Ly\alpha$ coupling case suffers from this uncertainty, although our
main trends and conclusions will not change.  Finally, in the strong heating case, the influence of
resolution should be negligible.  This is natural since in those two models, the dominant 
signal is from the neutral gas.

\item To derive the mass evolution of MHs undergoing photoevaporation we use the fitting formula 
by \cite {Iliev2005}. 
The formula reproduces the results from their numerical simulation, but it is approximate. Moreover,
it is accurate only for a constant ionization flux, while in the simulations by C06, the flux 
assumes different values at different times and locations. We tried two different methods to 
derive eq.~(\ref{mt}) in the presence of  a time dependent flux, corresponding to 
eq.~(\ref{mi}) and eq.~(\ref{dm}),
which consistently predict the $\delta T_b$ within few percent.
As in C06, for most halos, the evaporation time is comparable or smaller than the 
time between two outputs (for example, in the extreme suppression model $\approx$ 66\% of the 
cells lose 80\% of their MH mass between two outputs). We also do not expect the use of those 
formulae to introduce uncertainties in addition to those already present in the C06 implementation.

\item To calculate the 21~cm flux from an individual minihalo we use the formula from 
\cite {Iliev2002} and then we integrate over the mass function in eq.~(\ref{newmf}) to
obtain the beam-averaged effective differential brightness temperature of eq.~(\ref{dT_mh}). 
This formula is only suitable for halos that are in post-collapse equilibrium 
with a TIS profile. Yet, when a halo is being photoevaporated, both its density
and temperature profile may be substantially different\footnote{And in general, also if
undisturbed, the halo could follow a different shape (e.g. \citealt {Navarro1997}).}.
The new profile should depend on the local ionization flux. If we define the dynamical timescale as 
$t_{\rm dyn}=\pi r_{TIS}/v_c$, where $v_c$ is the halo circular velocity, for a halo with mass 
$10^7$~M$_\odot$ at redshift 15, $t_{\rm dyn} \approx 150$ Myr while $t_{\rm ev} \approx 230$ Myr 
if $F_0=1$. So the timescales are comparable and the halo profile could be changed
during photoevaporation. In fact, also \cite {Iliev2005} notice that the typical evaporation
time is long enough to see a gas-dynamic back-reaction. Since the effects of this change on 
the 21~cm signal of MHs is still not clear, we ignore the effect in this work. 
In addition, heating of the IGM can also 
change the profile of a halo, make it less concentrated and thus 
increase the 21~cm signal \citep{Oh2003a}.
On the other hand, the 
equation $T_s \gg T_{\rm CMB}$ will always be satisfied, so the impact of the temperature change
on the 21~cm signal should be much less than that of the change on the density profile.

\end{itemize}

\section{CONCLUSIONS}\label{con}

We have studied the 21~cm signal from both the IGM and minihalos based on the 
reionization simulations run by \citet{Ciardi2006}. These include a ``sub-grid'' 
prescription to take into 
account the photoevaporation of MHs in two different models for their formation 
(extreme suppression and reformation).
Our estimates follow the changes in the mass function of MHs and consider
three heating case of the IGM. We find that:
\begin{itemize}

\item the 21~cm signal tightly
relates to the thermal history of the IGM. In all the three configurations
(heating purely by the UV photons, Ly$\alpha$ and UV heating, and strong heating),
the IGM dominates the 21~cm signal at almost all redshifts, except at a very
early stage when the gas is still cold and a Ly$\alpha$ background has not been constructed 
or at the end of reionization in the reformation
model, when almost all the remaining neutral gas is in MHs. 
The small contribution of minihalos to the total 21~cm flux is partially due to the fact 
that the collapsed fraction of MHs is relatively small at high redshift.
Minihalos are significant emission sources in the absence of IGM heating,
however, in partially ionized regions which are heated by UV photons, where 
there is enough residual neutral gas and the temperature
is much larger than $T_{\rm CMB}$, the IGM also provides 
a contribution to the emission signal. 
In the presence of a Ly$\alpha$ background, neutral gas will be visible in strong 
absorption until a later stage when neutral regions are rare.  
In such case, totally neutral regions contribute more than partially ionized regions 
and MHs to the total 21~cm signal.

\item Although MHs are directly visible in 21~cm signal only under particular circumstances, they
nevertheless play a significant role in affecting the ionization state of the IGM and the 
corresponding 21~cm flux. Their presence can  delay reionization by as much as ${\Delta}z\approx 4$,
depending on the effect of feedback on their formation (C06).  More specifically, in the pure
UV heating case,  the 21~cm signal from the IGM in the reformation model is weaker than that in 
extreme suppression model at the same redshift, since MHs delay the photoheating associated with
reionization. In the strong heating case on the other hand, the 21~cm emission is higher in the 
reformation model than in the extreme suppression model at the same redshift, because absorption 
from MHs leaves more neutral gas in the IGM. This difference shows that in the pure UV 
heating case, the temperature is
the most important factor which can impact the ${\delta}T_b$ of the IGM, while in the strong heating case
the most important factor is $n_{HI}$. 

\item While the 21 cm signal from MHs exhibits differences from the IGM signal,
including the pixel distribution, detecting such features will not be within the capabilities of the planned
generation of radio telescopes (e.g. LOFAR\footnote{http://www.lofar.org/}, 
MWA\footnote{http://www.haystack.mit.edu/mwa/index.html}, 21cmA\footnote{http://21cma.bao.ac.cn/}).
In principle, MHs are more abundant in rich cluster
regions, which implies a higher 21~cm signal. In practice, in these regions
 also contain more  ionizing sources that suppress
MHs formation and evolution, reducing their signal.
\end{itemize}

Conclusions similar to those derived from our simulations have been reached by previous analytic estimates
(e.g. \citealt{Oh2003,Furlanetto2006a}).
For example, \cite {Furlanetto2006a} calculated the signal from both the IGM and the minihalos, using
a semi-analytic approach that includes the X-ray heating and feedback effects on halo formation. 
They found that in the presence of Wouthuysen-Field coupling, MHs are always buried by the IGM
signal, even if
the Ly$\alpha$ background is small. On the other hand, \citet{Shapiro2006} concluded that
before reionization but after $z\approx 20$, ignoring the radiative processes but taking 
into account shock heating, minihalos could dominate the total signal.
This conclusion is consistent with our pure UV heating case at the early stage.

\section{ACKNOWLEDGMENTS}

We thank Ilian T. Iliev, Jordi Miralda-Escude, and Yidong Xu for helpful discussions.  
ES thanks the Max Planck Institute for Astrophysics for their hospitality.
XC is supported by the NSFC Distinguished Young Scholar Grant No.10525314, Key Grant No. 10503010,
by the CAS grant No. KJCX3-SYW-N2, and by the MoST 973 project grant No.2007CB815491.

{}
\end{document}